 \documentclass[preprint2]{aastex}

\newcommand{\kms}{km\,s$^{-1}$}

\newcommand{\ergs}{erg\,s$^{-1}$}

\newcommand{\msun}{M$_{\odot}$}

\newcommand{\heb}{He~{\sc ii}}
\newcommand{\sid}{Si~{\sc iv}}

\shorttitle{HD~93250 resolved by the VLT interferometer}
\shortauthors{Sana et al.}

\begin{document}


\title{The non-thermal radio emitter HD 93250 resolved by long baseline interferometry\altaffilmark{1}}


\author{H. Sana}
\affil{Astronomical Institute 'Anton Pannekoek', University of Amsterdam, Postbus 94249, 1090 GE, Amsterdam, The Netherlands}
\email{h.sana@uva.nl}
\author{J.-B. Le Bouquin}
\affil{UJF-Grenoble 1 / CNRS-INSU, Institut de Plan{\'e}tologie et d'Astrophysique de Grenoble (IPAG) UMR 5274, Grenoble, France}
\author{M. De Becker}
\affil{Institut d'Astrophysique et G\'eophysique, Li\`ege University, All\'ee du 6 Ao\^ut 17, B-4000 Li\`ege, Belgium}
\author{J.-P. Berger\altaffilmark{2}}
\affil{European Organisation for Astronomical Research in the Southern Hemisphere (ESO), Casilla 19001, Santiago 19, Chile}
\author{A. de Koter\altaffilmark{3}}
\affil{Astronomical Institute 'Anton Pannekoek', University of Amsterdam, Postbus 94249, 1090 GE, Amsterdam, The Netherlands}
\and
\author{A. M\'erand}
\affil{European Organisation for Astronomical Research in the Southern Hemisphere (ESO), Casilla 19001, Santiago 19, Chile}


\altaffiltext{1}{Based on observations collected under program IDs 086.D-0586 and 386.D-0198 at the European Organisation for Astronomical Research in the Southern Hemisphere, Chile}
\altaffiltext{2}{UJF-Grenoble 1 / CNRS-INSU, Institut de Plan{\'e}tologie et d'Astrophysique de Grenoble (IPAG) UMR 5274, Grenoble, France}
\altaffiltext{3}{Astronomical Institute, Utrecht University, Princetonplein 5, 3584 CC, Utrecht, The Netherlands}



\begin{abstract}
As the brightest O-type X-ray source in the Carina nebula, HD~93250 (O4~III(fc)) is X-ray overluminous for its spectral type and has an unusually hard X-ray spectrum. Two different scenarios have been invoked to explain its X-ray properties: wind-wind interaction and magnetic wind confinement. Yet, HD~93250 shows absolutely constant radial velocities over time scales of years suggesting either a single star, a binary system seen pole-one or a very long period and/or highly eccentric system. Using the ESO Very Large Telescope Interferometer, we resolved HD~93250 as a close pair with similar components. We measured a near-infrared flux ratio of $0.8\pm0.1$ and a separation of $1.5\pm0.2 \times 10^{-3}$ arcsec. At the distance of Carina, this corresponds to a projected physical distance of 3.5~A.U. While a quantitative investigation would require a full characterization of the orbit, the binary nature of HD~93250 allows us to qualitatively explain both its X-ray flux and hardness and its non-thermal radio emission in the framework of a colliding wind scenario. We also discuss various observational biases. We show that, due to line-blending of two similar spectral components, HD~93250 could have a period as short as one to several years despite the lack of measurable radial velocity variations. 
\end{abstract}





\section{Introduction}



Massive stars of spectral-type O are typically soft and bright X-ray sources. Their X-ray emission is usually attributed to shocks in the dense region of the winds that grow from the intrinsic instabilities of the line driving mechanism. In the presence of a relatively strong magnetic field, the structure of the wind might be controlled by magnetic force lines and, at least in case of a bipolar field, the wind might be confined towards the equator, producing a typically harder X-ray emission compared to that expected from embedded wind shocks. In binary systems, an additional X-ray emission can also be generated by the supersonic collision of the two winds. One of the most comprehensive surveys of the X-ray properties of massive stars has recently been completed in the framework of the Chandra Carina Complex Project \citep{TBC11}, that has provided X-ray spectroscopy of 68 O-type stars in the Carina nebula. Yet not all sources are equally-well understood. 

Located at 7.5\arcmin\ NNW from Eta Carinae in the Trumpler 16 complex, \objectname[HD 93250]{HD~93250} \citep[O4~III(fc);][]{WSM10} is the brightest X-ray O star in Carina and has puzzled astronomers for over two decades. Assuming that it belongs to the Carina association, HD~93250 is too bright for its spectral type. Its X-ray luminosity, $L_\mathrm{X} \approx 1.5 \times 10^{33}$\ergs\ \citep{GFS11}, is a factor two to three larger than expected from the canonical $L_\mathrm{X}-L_\mathrm{bol}$ relation for O-type stars \citep{SRN06, NBO11}. Its X-ray emission is variable at the $\sim$20-50\%\ level on a time scale of months \citep{RNF09, GFS11} and its X-ray spectrum is extremely hard \citep[$\mathrm{k}T\sim2.3$~keV,][]{GFS11}. HD~93250 is also a known non-thermal (NT) radio emitter \citep{LCK95}.

Most of the puzzling properties of HD~93250 could be explained as part of a binary scenario where the collision between the wind of the two stars produces both an X-ray excess \citep{LMM90} and an NT radio emission component in the form of synchrotron radiation in the presence of a population of relativistic electrons \citep{DeB07}. The latter point is especially important in the sense that it provides the opportunity to investigate particle acceleration processes in massive star environments, in addition to evolved objects such as supernova remnants or binaries containing a compact object. For the wind collision scenario to produce enough X-ray and synchrotron radio flux, the winds from both components need to be of similar strength, and the companion is thus expected to be a bright O star. Yet, both high angular resolution observations \citep{NWW04} and RV campaigns \citep[e.g.,][]{RNF09} have failed to reveal a binary nature of the system. This has raised the question whether massive single stars could produce NT emission \citep{vLRB06}. As an alternative scenario, \citet{RNF09} suggested that the X-ray over-luminosity could originate from a magnetically confined wind but \citet{GFS11} noted that the strong non-thermal emission argued against such a scenario and called for further observations.

As discussed by \citet{SaE10}, long base-line interferometry is ideally suited to identify binaries in a separation regime that cannot be reached by either spectroscopy, typically limited to semi-major axes up to a few A.U., nor by other high resolution techniques, typically limited to a minimum angular separation of 20 mas, corresponding to $\sim$40 A.U.\ at the distance of the Carina complex. 

In this context, we observed HD~93250 using the Very Large Telescope Interferometer (VLTI) in an attempt to prove the binary nature of the system (Sect.~\ref{sect: obs}) and to perform a first characterization of the physical size of the orbit (Sect.~\ref{sect: discuss}).





\section{Interferometric measurements}   \label{sect: obs} 

\subsection{Observations and data reduction}
We observed HD~93250 at the ESO/VLTI \citep{HAB10}  with  the   Astronomical Multi-BEam combineR \citep[AMBER,][]{PMW07} on December 27 2010. Observations have been made in configuration A0-K0-G1, providing projected baselines ranging approximately from 50~m to 120~m. We recorded  spectrally dispersed fringes between $1.55-2.4\,\mu$m at a spectral resolving power ($R$) of 40. 

At VLTI the fast telescope guiding (STRAP) is performed in the V-band. For stars fainter than $V\sim11$~mag, the quality of the guiding is degraded and the amount of flux injected into the instrument is dramatically reduced. The offered limiting magnitude of AMBER ($H=5.5$~mag) has been measured on late-type stars, for which $H\sim6$~mag typically corresponds to $V\sim11$~mag. The offered magnitude thus results from the combination of the low flux limit in the observation band with the poor performance of the guiding for stars faint in $V$.

Early-type stars are intrinsically 0.7-0.8 mag brighter in the $V$-band compared to in the $H$-band \citep{MaP06} although the apparent $V-H$ color depends on the extinction properties. For stars as bright as HD~93250 in the $V$-band ($V=7.5$~mag) the telescope guiding performs very well. The fraction of flux injected into AMBER is optimal and the sensitivity of the observations is only limited by the intrinsic flux of the object in the observation band. As a consequence, HD~93250 ($H=6.7$) could be observed with the 1.8m Auxiliary Telescopes (ATs) despite the fact it is significantly fainter than the offered limiting magnitude.

Calibrated visibility and closure phase values were computed using the latest public version of the \texttt{amdlib} package \citep[v3.1,][]{TMC07,CUD09} and the \texttt{yorick} interface provided by the Jean-Marie Mariotti Center\footnote{JMMC software: \texttt{http://www.jmmc.fr}}. The main parameters of this pipeline are the criteria and the threshold used to reject individual frames. Based on our experience, we kept the 50\% best frames sorted by signal-to-noise ratio. We also discarded the fringes recorded with an optical path difference larger than 20~$\mu$m. For each spectral channel, we estimated the transfer function for square visibilities and closure-phase by averaging the measurements done on the nearest calibration star. This transfer function was divided and removed from the visibilities and closure-phases obtained for the scientific target. The dispersion between the consecutive points obtained for each observation of a calibration star is an estimate of the transfer function instability. This instability was quadratically added to the uncertainty on the transfer function. It is the dominant uncertainty term for visibilities, but is negligible for closure phase. The final uncertainty on the transfer function was propagated to the uncertainties on the calibrated quantities obtained for scientific targets. Note that these procedures are now fully embedded in the \texttt{amdlib} package. Concerning the wavelength calibration, we estimate the accuracy of our wavelength table to be 0.05~$\mu$m (that is about $4$\%) by cross-correlating our observed spectra with an atmospheric model as detailed in \citet[][]{KWB09}.


\begin{figure*}
  \centering
  \includegraphics[scale=0.7,angle=-90]{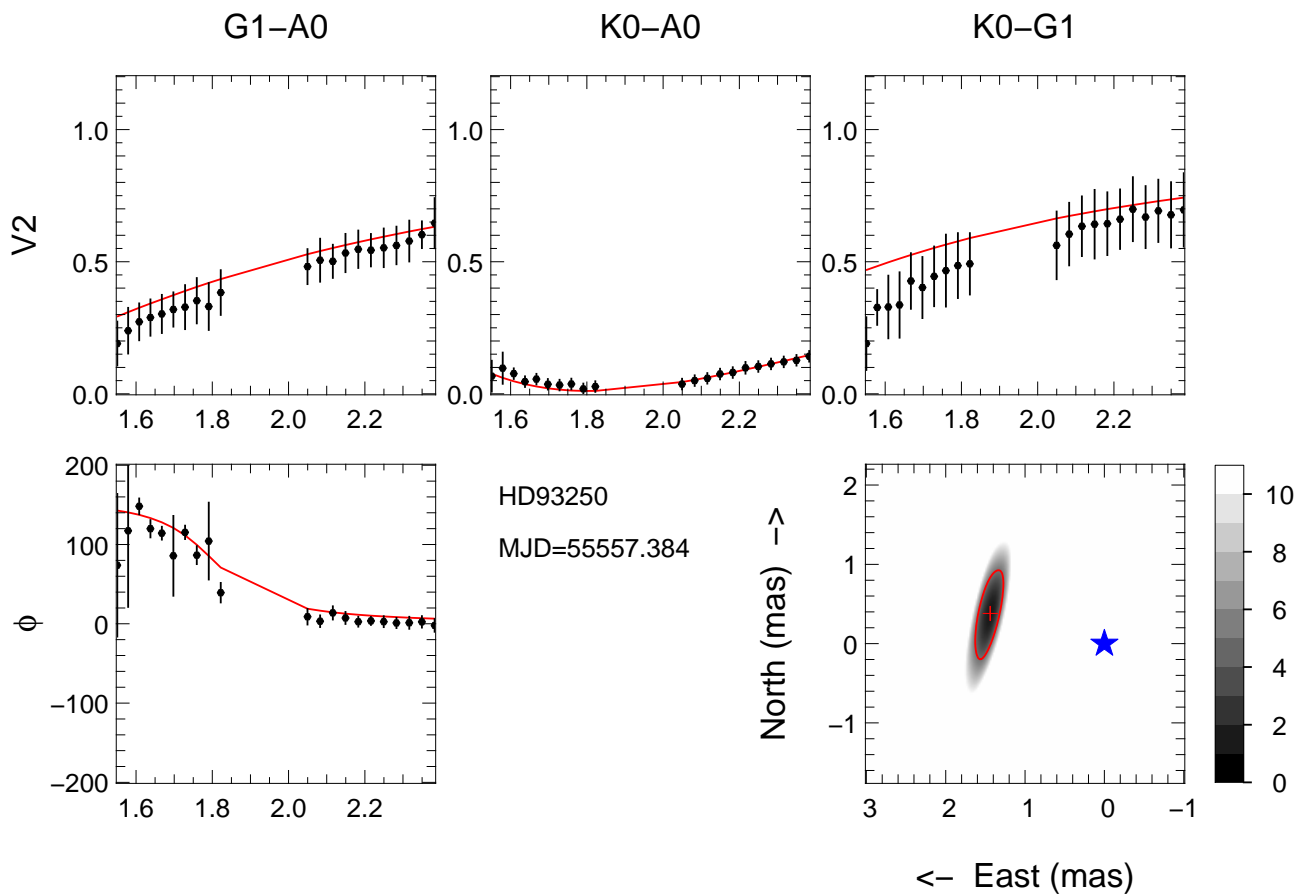}
  \caption{Square visibilities (top panels) and closure phase (bottom left panel) from the AMBER observations of HD~93250 compared with the best binary model. The bottom right panel provides the  $\chi^2_\mathrm{red}$ map with the flux separation of the best binary model. The best value is $\chi^2_\mathrm{red}\approx 0.9$. The red contour corresponds to $\chi^2=4.5$.} \label{fig: fit1}
\end{figure*}


\subsection{Modeling}
\label{sect: models}

The absolutely calibrated visibilities and closure phases are reported in Fig.~\ref{fig: fit1}. The target is well resolved in one baseline (K0-A0) and significantly less resolved on the two others (G1-A0 and K0-G1). The closure phase shows a clear departure from 0 (or 180) degrees. As expected, it is impossible to fit the data with a single-disk model as this yields an unrealistic diameter of 3.1~mas, a final $\chi^2_\mathrm{red} \approx 10$ and systematic discrepancies in the fit. Altogether, the binary nature of HD~93250 is conclusively revealed by these observations. 

We fitted the complete set of data (visibilities and closure phases) with a model composed of two unresolved stars with constant flux ratio over the H and the K bands, a valid assumption for O+O binaries. The separation, the orientation and the flux ratio were the only free parameters of the fit. We then explored the full parameter space to find the best $\chi^2$ value and we obtained a very good fit with  $\chi^2_\mathrm{red}\approx 0.9$, as shown in Fig.~\ref{fig: fit1}. Thanks to the combinations of three baselines of different orientations and of the H and the K bands, the position angle of the binary is unambiguous. By chance, an additional ``locking point'' for the binary model is provided by the visibility minimum observed in baseline K0-A0.

The best fit results are summarized in Table~\ref{tab: target}.  The exact shape of the astrometric error ellipse can be seen in the $\chi^2_\mathrm{red}$ plane of Fig.~\ref{fig: fit1}. In Table~\ref{tab: target}, $\rho$ is the projected separation; $\theta$, the position angle measured East from North; and $L_\mathrm{2,K}/L_\mathrm{1,K}$, the $K$-band flux ratio.

\subsection{Additional observations}
HD~93250 has been observed two more times with VLTI/AMBER. On March 24 2011, we used the 8.2-m telescopes (UTs) in configuration UT1-UT3-UT4 but we could not resolve the binary. This is in agreement with the very small angular separation between the components and the smaller baseline provided by the UTs. On March 29 2011, we used the ATs and we resolved HD~93250 again. We measured a binary separation of $\mathrm{dE}=1.37\pm0.25$~mas and $\mathrm{dN}=0.52\pm0.75$~mas and a flux ratio of $0.85\pm0.10$, which is fully consistent with our observation of December 2010 reported in Table~\ref{tab: target}. Given the uncertainties, a rotation on the plane of the sky smaller than about 30\degr\ (thus $1/12$ of an orbital cycle) would remain within the error bars. This second observation thus suggests a period larger than 3 yr, a high inclination or a very eccentric system. Alternatively, we might have observed HD~93250 at the same orbital phase ($\pm$0.5, as the sign of the closure phase is unconstrained in the observation of March 2011).


\begin{table}
\centering
\caption{Best-fit parameters of the binary model to the VLTI observations. }
\label{tab: target}
\begin{tabular}{lrrrrrrrrr}
\hline 
Parameter    & Dec. 2010 & Mar. 2011 \\
\hline
 dE (mas)                     & $1.44\pm 0.20$  & $1.37\pm 0.25$  \\
 dN (mas)                     & $0.38\pm 0.45$  & $0.52\pm 0.75$  \\
 $\rho$ (mas)                 & $1.49\pm 0.22$  & $1.47\pm 0.35$  \\
 $\theta$ (\degr)             & $  76\pm 17  $  & $  69\pm 27  $  \\
 $L_{2,\mathrm{K}}/L_{1,\mathrm{K}}$ & $0.81\pm 0.10$  & $0.85\pm 0.10$  \\
\hline
\end{tabular}
\vspace{0.cm}
\flushleft
\end{table}

\section{Discussion} \label{sect: discuss}

\subsection{Optical observations} \label{sect: optical}
\citet{RNF09} measure RVs from six optical spectra of HD~93250, obtained from 1999 to 2004, and concluded that variations are absent, with a measured 1-$\sigma$ dispersion of 1.3~\kms. We reanalyzed the  \citeauthor{RNF09} data set, complemented by two additional observations obtained with the Fiberfed Extended Range Optical Spectrograph (FEROS) and available from the ESO archives. The epochs of these supplementary observations are HJD - 2,450,000 = 3365.801 (PI: Bouret) and 3738.828 (PI: Casassus), i.e., respectively, nine months and almost two years after the last \citeauthor{RNF09} RV measurements. Despite the high signal-to-noise ratio ($S/N$) of the data (typically about 300), we could not detect any RV variation either. The maximum difference between the RVs from the different epochs is 3.6~\kms\ and the 1-$\sigma$ dispersion around the mean is 1.1~\kms. We also searched for line profile variations in the \heb\ and \sid\ lines, with a negative result. The relative peak-to-peak variations of the amplitude and of the full width at half-maximum (FWHM) of e.g., the \heb\,$\lambda$4541 line, are below  5 and 10\%\, respectively, pointing again to an excellent stability of the HD~93250 spectrum over the years.

\subsection{Physical constraints}
HD~93250 belongs to Tr~16 in the Carina association. Adopting a distance of $2.35$~kpc \citep{Smi06}, the angular separation translates to a projected physical distance of 3.5~A.U. The second component appears of a quite similar brightness and is thus likely an O star as well. We estimate the mass-ratio using a mass-luminosity relation given by $L_\mathrm{bol}\sim M^{2.04}$ \citep{BdMC11}. Considering no color difference between the two components of the pair, thus no difference in their bolometric correction, we derive a mass ratio $M_2/M_1$ of 0.9. 

Assuming a circular orbit and a primary mass $M_1$ of 47~\msun\ as appropriate for an O4~III star \citep{MSH05}, a 3.5 A.U.\ semi-major axis corresponds to an orbital period of about 250~d and an orbital velocity of 70~\kms. A maximum inclination of 2.5\degr\ is required to hide such a signal at the 5-$\sigma$ level in the noise associated the RV measurements. Assuming random orientation of binary orbits in space, one expects that such a low inclination is encountered with a 1:10,000 frequency and is thus very rare.


\begin{figure}
  \centering
  \includegraphics[width=\columnwidth]{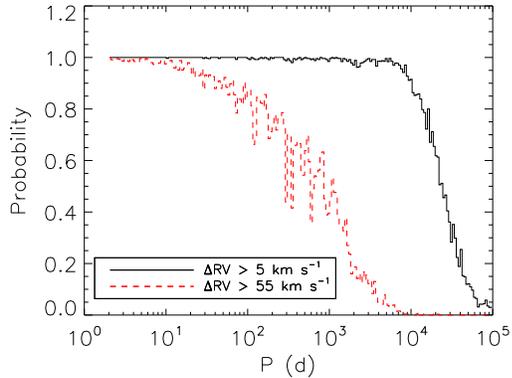}
  \caption{Probability that a RV shift larger than, respectively, 5 and 55 \kms\ would occur in between any of the spectroscopic observations discussed in Sect.~\ref{sect: optical} as a function of the orbital period $P$.} \label{fig: proba}
\end{figure}



\begin{figure}
  \centering
  \includegraphics[width=\columnwidth]{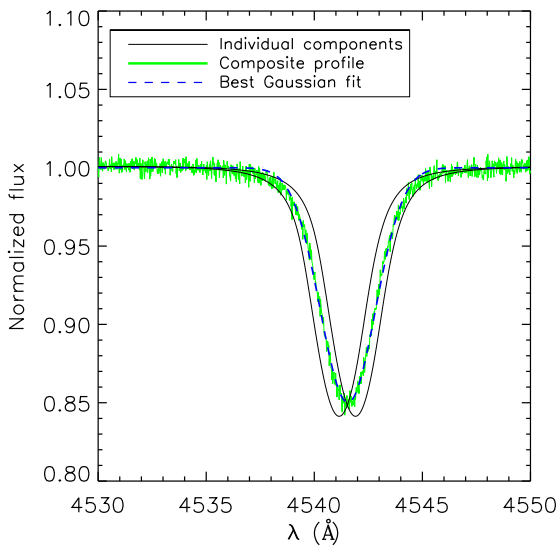}
  \caption{Normalized composite \heb~$\lambda$4541 line profile in an O4~III+O4~III binary with a separation of 50~\kms\ between the two components (plain lines) together with the best-fit Gaussian profile (dashed line) to the line profile. Gaussian noise corresponding to a $S/N$ of 300 has been added to the composite profile before fitting.} \label{fig: spec}
\end{figure}

Alternatively, HD~93250 might be a long period, highly eccentric system. We used the method described in \citet{SGE09} to estimate the likelihood to observe a significant RV shift as a function of the orbital period given the sampling provided by the optical observations and using no constraint on the orbital inclination. The probability that HD~93250 would have displayed a RV shift larger than 5~\kms\ is very close to 1.0 up to  periods of $10^4$~d, then drops sharply for longer periods (Fig.~\ref{fig: proba}). As no such a shift is seen, this suggests a minimum orbital period of $\sim$30~yr, which would imply an eccentricity as large as 0.9 to match the current separation.

\begin{figure}
  \centering
  \includegraphics[width=\columnwidth]{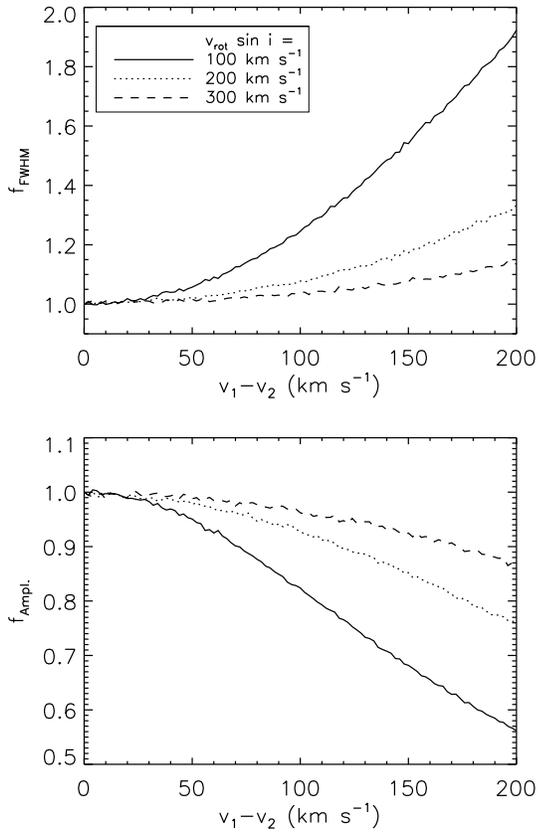}
  \caption{Relative variation of the FWHM ($f_\mathrm{FWHM}$) and of the amplitude ($f_\mathrm{Ampl}$) of the \heb~$\lambda$4541 line in an O4~III+O4~III binary as a function of the RV separation $v_1-v_2$ between the two components. Three rotational broadenings are considered.} \label{fig: fwhm}
\end{figure}

There is however an additional detection bias that results from the blending of two equal-strength lines, which is not taken into account in \citet{SGE09}. Indeed the RV measurement obtained when applying single-Gaussian fitting to blended lines will remain constant (Fig.~\ref{fig: spec}) until evidence for a double-lined profile is seen (typically at $\Delta$RV $>$ 100~\kms). To estimate this bias, we have generated synthetic profiles of O4~III+O4~III binaries using the non-LTE atmosphere code FASTWIND \citep{PUV05}, for various RV shifts up to 200~\kms\ and for different rotational broadening. Accounting for the typical $S/N$ of our optical data, we have then fitted a single Gaussian profile. As expected, the measured RVs show no significant variations. For a typical rotational broadening of 100~\kms, as appropriate for HD~93250, the measured amplitudes and FWHMs are changing by less that 5 and 10\%\ respectively for a RV difference between the two components of less than 55 and 70~\kms\ (Fig.~\ref{fig: fwhm}). Using these new values for the detection thresholds, we note that the detection probability already starts to drop for period larger than a few 100s days. 

As a conclusion, the fact that RV campaigns have failed to reveal the binary nature of HD~93250 does not necessary imply exceptionally long periods, high eccentricities and/or small inclination angles. It might also result from observational biases affecting binaries with (quasi) identical components.


\subsection{Colliding winds} \label{cwb}

Using the formalism of \citet{Uso92} adapted to O+O binaries, we estimate the X-ray luminosity that would be produced by the collision of two O4~III winds of equal strength to be of the order of $10^{33}$~\ergs. There is enough uncertainty in the mass-loss rate associated with HD~93250 \citep[see the discussion in][]{GFS11} to accommodate the range of periods and separations discussed in the previous section. It is thus not possible to use the X-ray luminosity to further constrain the orbital properties.  Assuming that the wind-wind collision is adiabatic, as expected for wide binaries, the X-ray flux produced in the wind collision should scale with the separation $D$ as $L_\mathrm{X}\sim D^{-1}$ \citep{SBP92}. Given that \citet{RNF09} reported a variation of the X-ray flux of the order of 40\%, one would thus expect HD~93250 to have a minimum eccentricity of 0.2.

In a scenario where the observed radio synchrotron radiation is produced by colliding winds, the wind-wind interaction region should be located out of the bulk of the stellar wind material. Otherwise, the non-thermal radio photons suffer from significant free-free absorption and cannot escape the radio photosphere. Adopting the mass loss rate and the terminal velocity given by \citet{MSH05ww} (respectively 5.62\,$\times$\,10$^{-7}$\,M$_\odot$\,yr$^{-1}$ and 3000 km\,s$^{-1}$), and using equations of \citet{WrB75} and \citet{LCK95}, we estimate the radius of the $\tau_\nu = 1$ photosphere using the same procedure as in the case of HD~168112 in \citet{Becker2004}. The two frequencies investigated by \citet{LCK95}, i.e. 8.64 and 4.80\,GHz, yield respectively photosphere radii of 1.4 and 2.2\,A.U. As the two components of the system are very similar, the wind-wind interaction region is expected to be located close to mid-separation, i.e. about 1.7-1.8 A.U.\ according to our results and assuming no significant inclination effect. Interestingly, these values are in good agreement with the detection of HD~93250 at 8.64\,GHz, and its non-detection at 4.80\,GHz \citep{LCK95}. We emphasize that these considerations are likely to be epoch-dependent. For instance, an increase of the stellar separation in an eccentric orbit could lead to the detection of the synchrotron radiation at 4.80\,GHz as well. This scenario can only be verified through a multi-frequency radio monitoring of HD~93250 on time scales covering a significant fraction of the orbital period.

\section{Conclusion}
Using the ESO/VLTI, we provided evidence that HD~93250 is an O+O binary system. The binary nature gives a natural explanation of many of the intriguing properties of the object. In particular, it can explain the X-ray over-luminosity and NT radio emission in the framework of a colliding wind scenario. The latter point lends further support to the so-called `standard scenario' where particle acceleration occurs in the wind-wind interaction of a massive binary system. Because we see no change in the orientation of the system in our two VLTI observations separated by three months, we conclude that HD~93250 has a likely orbital period of the order of one to several years. A significant eccentricity is also required to explain the variations observed in the X-ray flux. 

We further show that the non-detection of a RV signature based on eight high-resolution spectra covering a seven-year period of time does not require a particularly unfavorable orbital configuration (extremely low inclination, long period and/or high eccentricity). It can also be explained by observational biases resulting from line-blending in nearly equal-mass systems. Further spectroscopic and high-angular observations are require to characterize the orbit of the system, which is needed to quantitatively understand the X-ray and NT radio processes originating from the wind-wind collision.

Together with \objectname[HD 168112]{HD~168112}, HD~93250 was one of the only two known O-type NT radio emitters for which the binary scenario was questioned. Our results provide further evidence that binarity and NT radio emission go hand-in-hand. As was the case for HD~93250, RV campaigns have so far failed to confirm the binary nature of HD~168112. Long base-line interferometry might thus provide a complementary diagnostic tool to test the multiplicity properties of HD~168112 in particular and of single non-thermal radio emitters in general.

\acknowledgments
The authors warmly thank the VLTI team of the Paranal Observatory. CNRS is acknowledged for having supported this work with the allocation of Guaranteed Time Observations. We are also greatful to Nolan Walborn for interesting discussion. 



{\it Facilities:} \facility{VLTI (AMBER)}



\clearpage



\end{document}